\documentclass[aps,prd,twocolumn,superscriptaddress,showpacs,showkeys,
nofootinbib]{revtex4-1}

\usepackage{latexsym}
\usepackage{amsmath}
\usepackage{amssymb}
\usepackage{amsfonts}
\usepackage{multirow}
\usepackage{color}

\usepackage{supertabular}
\usepackage{placeins}
\usepackage{epsfig}
\usepackage{graphicx}

\begin{document}

\title{Production of single-charmed baryons in a quark model approach}


\author{D. Iglesias-Ferrero}
\affiliation{Departamento de Física Fundamental, Universidad de Salamanca, E-37008 Salamanca, Spain}

\author{D. R. Entem}
\email[]{entem@usal.es}
\affiliation{Instituto Universitario de F\'isica
Fundamental y Matem\'aticas (IUFFyM), Universidad de Salamanca, E-37008 Salamanca, Spain}
\affiliation{Grupo de F\'isica Nuclear, Universidad de Salamanca, E-37008 Salamanca, Spain}

\author{F. Fern\'andez}
\email[]{fdz@usal.es}
\affiliation{Instituto Universitario de F\'isica
Fundamental y Matem\'aticas (IUFFyM), Universidad de Salamanca, E-37008 Salamanca, Spain}
\affiliation{Grupo de F\'isica Nuclear, Universidad de Salamanca, E-37008
Salamanca, Spain}

\author{P. G. Ortega}
\email[]{pgortega@usal.es}
\affiliation{Departamento de Física Fundamental, Universidad de Salamanca, E-37008 Salamanca, Spain}
\affiliation{Instituto Universitario de F\'isica
Fundamental y Matem\'aticas (IUFFyM), Universidad de Salamanca, E-37008 Salamanca, Spain}

\date{\today}

\begin{abstract}

The production of single-charmed baryons $\Lambda_c^+ \Lambda_c^-$, $\Lambda_c^+ \Sigma_c^-$+h.c. and $\Sigma_c^+ \Sigma_c^-$ in $p\bar p$ collisions
is studied in the framework of a constituent quark model which has satisfactorily described the $N\bar N$ system and the strangeness production $p\bar p\to \Lambda\bar \Lambda$, $\Lambda\bar\Sigma$ and $\Sigma\bar\Sigma$ processes. Predictions on the total cross sections are analyzed for different approaches to the underlying $n\bar n\to c\bar c$ process, mediated by one gluon annihilation diagrams. The results indicate that the cross section is of the order of $1$ nb between $10-14$ GeV for $\Lambda_c^+\Lambda_c^-$ and $\Lambda_c^+ \Sigma_c^-$ channels, and around $0.01-0.1$ nb for $\Sigma_c^+\Sigma_c^-$.  This estimations can be relevant for their future search in facilities like $\overline{P}$ANDA.
\end{abstract}



\maketitle



\section{Introduction}

Charmed-hadron physics has been on the spotlight in the last years with the observation of many new excited states of charmed baryons and mesons, as well as exotic states with four or five minimum quark content (e.g., see a review on Ref.~\cite{Wiedner:2011mf,RevModPhys.90.015004}).
The construction of the $\overline{P}$ANDA experiment~\cite{PANDA:2009yku} at the antiproton facility FAIR at GSI laboratory plans to push forward the research on charmed hadron spectroscopy and reactions with high-accuracy measurements~\cite{Wiedner:2011mf}.

Regarding charmed baryons, the program contains not only a study of the spectroscopy, but also in-medium effects, charm-anticharm mixing, the search for CP violation or studies of precise measurements of $\Lambda_c$ decays. A required previous step is the estimation of charm-production cross sections in $p\bar p$ collisions. Indeed, charm-production processes in proton-antiproton reactions are important to test the mechanisms of charm quark pair creation, which is in the regime of nonperturbative QCD.
 In particular, increasing interests are being received to the exclusive production reactions such as $p\bar p\to\Lambda_c^+\Lambda_c^-$.

This process is the prototype reaction to analyze the underlying process of charm quark pair creation in baryons.
The process recalls the hyperon production reaction $p\bar p\to \Lambda\bar\Lambda$.
Traditionally, two different pictures have been used to describe the strangeness production. One of them is based on the t-channel meson exchange~\cite{haidenbauer1992reaction}, whereas the second works at the quark level being the strangeness created by $q\bar q$ annihilation and subsequent $s\bar s$ creation by s-channel gluon exchanges~\cite{Ortega:2011zza}.

Both models can be extended to the $Y_c\bar Y_c$ production. The extension of the meson exchange model to the charm sector is based on $SU(4)$ flavor symmetry. Accordingly, the elementary
charm production process is described by t-channel $D$ and $D^*$ meson exchanges. Such a model has been applied to the charm production in Ref.~\cite{Haidenbauer:2016pva}. However, up to date, the quark model has not been extended to the charm production.

Other models, included quark-gluon string model~\cite{Khodjamirian:2011sp,Kaidalov:1994mda}, effective Lagrangian models~\cite{Lin:2021wrb,Shyam:2017gqp,Sangkhakrit:2020wyi} and perturbative QCD calculation within the handbag approach~\cite{Goritschnig:2009sq}, have been used to describe the charm production, with a disparity of predictions ranging several orders of magnitude.

One characteristic feature of the $Y_c\bar Y_c$ production
is that different final channels are produced via selected
isospin channels: the $\Lambda^+_c \bar \Lambda^-_c$ is produced via the isospin
singlet ($I=0$) channel, whereas the $\Lambda^+_c \bar \Sigma ^-_c$  +h.c. goes
through the isospin triplet ($I=1$). This fact can be used to differentiate models which involves different physics.
In the conventional meson-exchange picture, the charm
production transition interaction is weaker than the $Y\bar Y$
production mediated by kaon exchanges, but it predicts
larger cross sections than alternative models based on s-channel
exchanges. However this latter models will show a dominance of
spin-triplet channels, since the spin of the $\Lambda_c$ is entirely
carried by the c-quark, so a triplet $c\bar c$, produced by an
effective vector as the gluon, guarantees a triplet $\Lambda_c \bar \Lambda_c$ final state.

Unfortunately, unlike the hyperon-antihyperon reactions, the charmed baryon
production lacks experimental data, so there is no way to
discriminate among models. Nevertheless, the theoretical
calculations allow to predict the order of magnitude of
the observables, which can be of help for the future
experimental search.

In view of the experimental and theoretical interests in such reactions, in this paper we analyze the $Y_c\bar Y_c$ production reaction using a widely-used constituent quark model~\cite{Vijande:2004he,Segovia:2013wma}.
The model has been extensively used to describe
hadronic spectroscopy and reactions~\cite{Garcilazo:2001md,Vijande:2004at,Fernandez:2019ses,Ortega:2012cx,Ortega:2016syt}, as well as the  $p\bar p$ cross section and positronium level shifts~\cite{Entem:2006dt} and the analogue strangeness process of proton-antiproton to hyperon-antihyperon~\cite{Ortega:2011zza,GarciaOrtega:2012lgi}.
Besides the one-gluon exchange~\cite{manohar1984chiral}, the constituent quark model encodes
constituent masses, emerging as a dynamical mass generated by the
spontaneous breaking of the $SU(3)_L\otimes SU(3)_R$ symmetry of the QCD Lagrangian.
Hence, as a consequence of Nambu-Goldstone theorem, eight massless bosons identified with the members
of the pseudoscalar octet $\{\pi, K, \eta\}$, corresponding with the eight broken
generators, are expected to be
exchange between the massive quarks. Thus, this model shares features of meson-exchange and quark-gluon models
in an unified framework.

In this approach, the $n\bar n\to c\bar c$ transition is described via a gluon annihilation diagram, building the hadronic $p\bar p\to Y_c\bar Y_c$ transition interactions by means of the Resonating Group method, which folds the quark level interaction with the wave functions of the baryons under play.
This  feature implies that the $\Lambda_c^+\Lambda_c^-$ production will occur in a pure spin-triplet state, being the spin-singlet channels completely suppressed.
Nambu-Goldstone boson exchanges do not affect the transition potential, but constrains the initial and final state interactions.

The paper is organized as follows: In Section~\ref{sec:model} we describe the details of the constituent quark model and the calculation of the cross section in a coupled-channels approach. In Sec.~\ref{sec:results} we present the results and a comparison with other theoretical estimations, and in Sec.~\ref{sec:summary} we give a short summary.

\section{The model}\label{sec:model}

The predictions for the production of $\Lambda_c$/$\Sigma_c$ are based on previous
analysis of strangeness production in proton-antiproton collisions~\cite{Ortega:2011zza,GarciaOrtega:2012lgi}.
Such calculations, and those in the present study, were done in the framework of the constituent
quark model described in Ref.~\cite{Entem:2006dt} for the $N\bar N$ system. All the parameters of the model have been
constrained from previous calculations of hadron phenomenology, so in that sense the results are parameter free.

Based in Diakonov's picture of the QCD vacuum as a dilute instanton liquid~\cite{Diakonov:2002fq},
the model assumes that the quark acquires a dynamical mass due to the interaction with fermionic zero modes
of individual instantons. The momentum-dependent mass vanishes for large momenta, and acts
as a natural cut-off of the theory. This scenario can be modelled with the chiral invariant simple Lagrangian~\cite{Diakonov:2002fq}

\begin{equation}\label{eq:diakon}
 \mathcal{L}=\bar{\psi}(i\gamma^\mu \partial_\mu-M U^{\gamma_5})\psi
\end{equation}
where $U^{\gamma_5}=exp(i\phi^a\lambda^a\gamma_5/f_\pi)$, $\phi^a$ denotes the
pseudoscalar fields $(\vec\pi,K_i,\eta_8)$ with $i=(1,\ldots,4)$, $\lambda^a$
the $SU(3)$ flavor matrices and with
$M$ the constituent quark mass.
The momentum dependence of the constituent quark mass can be derived from the theory,
though a simpler parametrization can be obtained as
$M(q^2)=m_qF(q^2)$, where $m_q\sim 300$ MeV and

\begin{equation}
 F(q^2)=\sqrt{\frac{\Lambda_\chi^2}{\Lambda_\chi^2+q^2}}
\end{equation}
being $\Lambda_\chi$ a cut-off that fixes the chiral symmetry-breaking scale.

Expanding the Nambu-Goldstone boson field matrix $U^{\gamma^5}$ from Eq.~\eqref{eq:diakon} we have

\begin{equation}
U^{\gamma_5}=1+\frac{i}{f_\pi}\gamma_5\lambda^a\phi^a-\frac{1}{2f_\pi^2}\phi^a\phi^a+\ldots
\end{equation}
where we can identify the constituent quark mass contribution in the first term.
Further terms give rise to quark-quark interactions mediated by boson exchanges.
Specifically, the second term represents the $\phi^a=\{\vec\pi,K_i,\eta_8\}$ one-boson exchange, with $i=(1,\ldots,4)$, whereas the third term illustrates a two-boson exchange, whose main contribution can be modelled as a scalar $\sigma$ exchange.

\begin{figure}[t]
\begin{center}
\includegraphics[width=.48\textwidth]{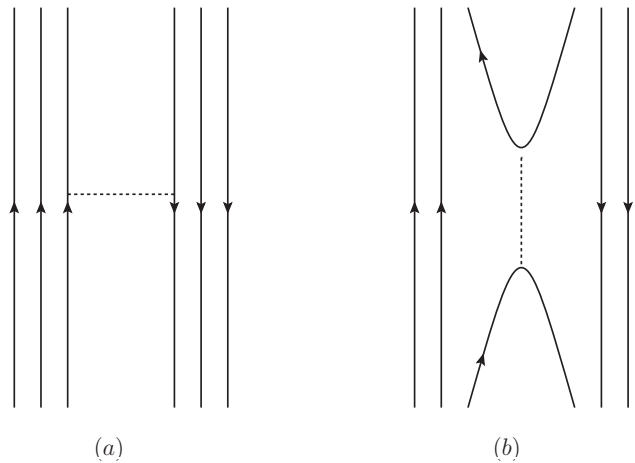}
\caption{\label{fig:NNb-int} Possible diagrams for the $N\bar N$ and $\Lambda_c
\bar \Lambda_c$ interactions.}
\end{center}
\end{figure}

Such boson exchanges, detailed in Refs.~\cite{Vijande:2004he,Ortega:2011zza}, allow to describe the $N\bar N$ and $Y_c\bar Y_c$ self-interactions, constrained by the light-quark potentials.
The basic $q\bar q$ potentials for such systems are

\begin{align} \label{eq:pots}
&V^{\pi}_{q\bar q} (\vec{q}) =
\frac{1}{(2\pi)^3} \,\frac{g_{ch}^2}{4m_q^2}
\,\frac{\Lambda^2_{\pi}}{\Lambda^2_{\pi}+q^2}
\,\frac{(\vec{\sigma}_i \cdot \vec{q})(\vec{\sigma}_j \cdot
\vec{q})}
{m_{\pi}^2 + q^2}
\,(\vec{\tau}_i \cdot \vec{\tau}_j),\nonumber
\\
&V^\eta_{q\bar q} =
-\frac{1}{(2\pi)^3}
\frac{g_{ch}^{2}}{4m_q^2}
\frac{\Lambda_\eta^2}{\Lambda_\eta^2+q^2}
\,\frac{(\vec{\sigma}_i \cdot \vec{q})(\vec{\sigma}_j \cdot
\vec{q})}{m_\eta^2+
q^2} (\cos \theta_P \lambda^8_i \, {\lambda^8_j}^\dagger-\sin \theta_P)  ,\nonumber
\\
&V^{\sigma}_{q\bar q} (\vec{q}) = -\frac{g_{ch}^2}{(2\pi)^3}
\,\frac{\Lambda^2_{\sigma}}{\Lambda^2_{\sigma}+q^2}
\,\frac{1}{m_\sigma^2 + q^2},
\end{align}
diagrammatically represented in Fig.~\ref{fig:NNb-int}a. The parameters of the model are presented in Table~\ref{tab:parametros}.
The chiral coupling constant $g_{ch}$ is determined from the $\pi NN$ coupling constant, through the relation,

\begin{equation*}
\frac{g^2_{ch}}{4\pi} = \left(\frac{3}{5}\right)^2 \frac{g_{\pi NN}^2}{4\pi}\frac{m_q^2}{m_N^2}
\end{equation*}
with $m_q$ the mass of the light $\{u,d\}$ quark~\cite{Vijande:2004he}.

\begin{table}
\begin{center}
\begin{tabular}{|c|cc|} \hline
\multirow{2}{*}{Quark Masses}   & $m_n$ (MeV) & 313 \\
   & $m_c$ (MeV) & 1763 \\ \hline
 & $m_\pi$ (MeV) & 138 \\
 & $m_\sigma$ (MeV) & 693 \\
 & $m_\eta$ (MeV) & 546.5 \\
Nambu-Goldstone Bosons& $\Lambda_\pi=\Lambda_\sigma$ (MeV) & 848.33 \\
 & $\Lambda_\eta$ (MeV) & 1025.96 \\
 & $g_{ch}^2/(4\pi)$ & 0.54 \\
 & $\theta_P$ ($^\circ$) & -15\\ \hline
  & $\alpha_0$ & 2.118 \\
 OGE & $\mu_0$ (MeV) & 36.976 \\
 & $\Lambda_0$ (fm$^{-1}$) & 0.113 \\ \hline
 \end{tabular}
\end{center}\caption{\label{tab:parametros} Quark-model parameters.}
\end{table}

Nevertheless, the charm production $n\bar n\to c\bar c$, with $n=\{u,d\}$, is beyond the chiral symmetry-breaking scale.
Hence, within our constituent quark model, such production mechanism is not mediated by Nambu-Goldstone boson exchanges but it is governed by QCD perturbative effects which are taken into account
through the one-gluon exchange term,

\begin{equation}
 \mathcal{L}_{gqq} = i\sqrt{4\pi\alpha_s}\bar{\psi}\gamma_\mu
G_c^\mu\lambda^c\psi,
\end{equation}
being $\lambda^c$ the $SU(3)$ color matrices and $G_c^\mu$ is the gluon
field. The strong coupling constant $\alpha_s$ has a scale dependence which allows to consistently describe light, strange and heavy mesons, whose explicit expression is,

\begin{align} \label{eq:alphascale}
 \alpha_s(\mu) = \frac{\alpha_0}{\ln\left(\frac{\mu^2+\mu_0^2}{\Lambda_0^2}\right)}
\end{align}
where $\mu$ is the reduced mass of the $q\bar q$ system and $\alpha_0$, $\mu_0$ and $\Lambda_0$ are parameters of the model (see Table~\ref{tab:parametros}).

Since quark-antiquark exchanges are not allowed in $N\bar N$ or $Y_c\bar Y_c$ systems, the
$N\bar N$ ($Y_c\bar Y_c$) interaction from one-gluon exchange only emerges as annihilation diagrams, that
is, in momentum representation~\cite{Entem:2006dt,Faessler:1982qt},

\begin{align}\label{eq:OGEani}
V^{A,OGE}_{q\bar q} (\vec{q}) =& \frac{\alpha_s}{8\pi^2m_{ij}}
\left( \frac{4}{9} - \frac{1}{12} \vec \lambda_1 \cdot \vec \lambda_2 \right)\nonumber\\ &
\left( \frac{3}{2} + \frac{1}{2} \vec \sigma_1 \cdot \vec \sigma_2 \right)
\left( \frac{1}{2} - \frac{1}{2} \vec \tau_1 \cdot \vec \tau_2 \right).
\end{align}
where $m_{ij}$ is the product of the masses of the interacting quarks. For the $N\bar N\to Y_c\bar Y_c$ those are $m_{ij}=m_n^2$, with $m_n$ the mass of the light quarks $\{u,d\}$. However, this choice may overestimate the contribution of the gluon annihilation interaction for a $n\bar n\to c\bar c$ process, whose energy is much larger than $m_n$. For that reason, we will compare the results with $m_{ij}=m_n^2$, which will be dubbed \emph{model (a)} for now on, with an averaged quark mass $m_{ij}=\overline{m_\Lambda}^2$ where $\overline{m_\Lambda}=\frac{1}{3}(2m_n+m_c)$, dubbed \emph{model (b)}. The value of the $\alpha_s(\mu)$ from Eq.~\eqref{eq:alphascale} will be scaled accordingly.

In the $q \bar q$ system there are also annihilation contributions from Nambu-Goldstone bosons (Fig.~\ref{fig:NNb-int}b) in the $N\bar N\to N\bar N$ and $Y_c\bar Y_c\to Y_c\bar Y_c$ reactions.
In our model, the real part of this potential can be obtained with a Fierz transformation of Eq.~\eqref{eq:pots}. Explicit expressions can be found in Ref.~\cite{Ortega:2011zza}.

The last ingredient of our constituent quark model is confinement, a non perturbative effect which prevents to have coloured hadrons. This potential does not contribute neither to the $N\bar N$ nor to the $Y_c\bar Y_c$ interactions, but allows to describe hadron spectroscopy and prevents the baryon collapse under the interactions described above.

\begin{figure}[t]
\includegraphics[width=.45\textwidth]{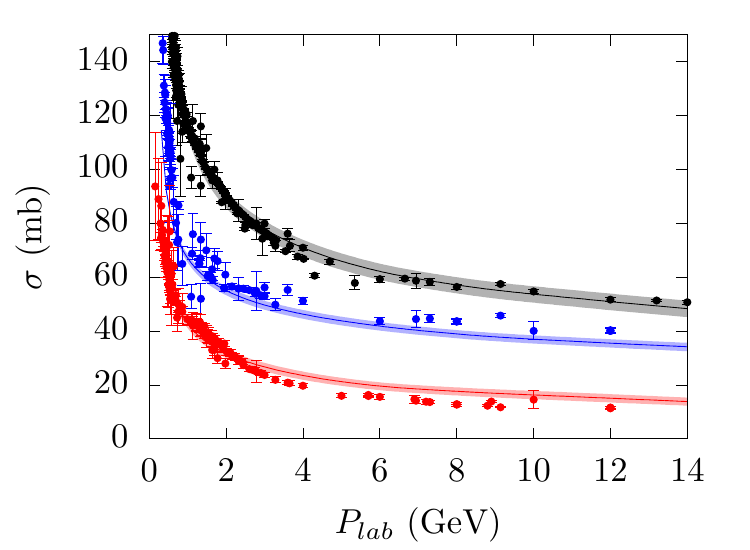}
\caption{\label{sigppbar} Total (black line), inelastic (blue) and elastic (red)
$p\bar p\to p\bar p$ cross sections. Experimental data from Refs.~\cite{ParticleDataGroup:2020ssz}. The shadowed bands around the lines show the sensitivity of the cross sections when the value of the $V_r$ and $W_i$ parameters are varied a $10\%$.}
\end{figure}

Finally, besides the interaction in the elastic channels, both $p\bar p$ and
$Y_c\bar Y_c$ annihilate into mesons, processes which are very complex to describe.
These contributions are usually parameterized in terms of optical potentials.
In order to evaluate the model predictions of the total cross section near the $Y_c\bar Y_c$ threshold, we will take the $Y_c\bar Y_c$ optical potential to be zero, as it is not expected to have a large influence in this energy region.
For the $N\bar N$ optical potential, with the aim of simplifying the description of the processes, we do not include spin-orbit or
tensor pieces. Thus, we take the form

\begin{equation} \label{ec:PotOptic}
 V_{\rm Opt}=\left(V_r+i\cdot W_i\right)e^{-\frac{d^2}{2}q^2},
\end{equation}
 where the  $d$, $V_r$ and $W_i$ parameters are fitted to the available total, elastic and inelastic $p\bar p\to p\bar p$ cross section up to $p_{\rm lab}=14$ GeV (see Fig.~\ref{sigppbar}), obtaining the values $d=0.509$ fm, $V_r=-0.184$ GeV$^{-2}$ and $W_i=-2.041$ GeV$^{-2}$. We consider a 10\% error in such parameters, though the impact of such uncertainty is negligible in the full $p\bar p\to Y_c\bar Y_c$ reaction.

The baryon-antibaryon interaction is obtained from the microscopic description using the
Resonating Group Method, obtaining an effective cluster-cluster interaction from the
underlying quark-quark dynamics, where the wave functions of the baryons act as natural
cut-offs of the $q\bar q$ potentials.

The wave function for the baryon (antibaryon) states are

\begin{align}
 \psi_B = \phi_B(\vec p_{\lambda},\vec p_{\rho})\chi_B \xi_c[1^3]
\end{align}
where $\chi_B$ is the spin-isospin wave function, $\xi_c$ is the color wave function and
$\phi_B$ is the orbital wave function, with $\vec p_\lambda$ and $\vec p_\rho$ the momenta of the $\lambda$ and $\rho$ modes. For the $N$ ($\bar N$) wave function, a good
description can be achieved with a Gaussian with range $b=0.518$ fm~\cite{Entem:2006dt}.

To solve the three-body Schr\"odinger
equation we use the Gaussian Expansion Method~\cite{Hiyama:2003cu}. In this method, the radial wave function is
expanded in terms of basis functions whose parameters are in geometrical progression,

\begin{align}
 \phi_B^{LM} = \sum_{n_\lambda,n_\rho}^{n_{\rm max}} C_{n_\lambda,n_\rho} \left[\phi_{n_\lambda,\ell_\lambda}(\vec p_\lambda)\phi_{n_\rho,\ell_\rho}(\vec p_\rho)\right]_{LM}
\end{align}
where $L$ is the total angular momentum, satisfying $L=\ell_\lambda\oplus\ell_\rho$, and $C_{n_\rho,n_\lambda}$ the coefficients of the base expansion. The $\phi$ functions are defined as

\begin{eqnarray}\label{eq:GEMwavef}
 \phi_{n,\ell m}(\vec p)&=& N_{n\ell}\, p^\ell\, e^{-\frac{1}{4\eta_n} p^2} Y_{\ell m}(\hat p).
\end{eqnarray}
with $N_{n\ell}$ the normalization of the Gaussian wave functions such that $\langle\phi_{n\ell}|\phi_{n\ell}\rangle=1$.

The coefficients $C_{n_\lambda,n_\rho}$ and the eigenenergies of the baryons are determined from the Rayleigh-Ritz variational principle,

\begin{align}
 \sum_{n=1}^{n_{\rm max}} \left[ (T_{n'n}^{\alpha'}-E N_{n'n}^{\alpha'})C_n^{\alpha'} +
 \sum_\alpha V_{n'n}^{\alpha'\alpha} C_n^\alpha= 0 \right]
\end{align}
with $n=\{n_\lambda,n_\rho\}$, $T_{n'n}^{\alpha'}$ and $N_{n'n}^{\alpha'}$ the kinetic and normalization operators, which are diagonal, and $V_{n'n}^{\alpha'\alpha}$ the underlying $q\bar q$ interaction from the constituent quark model detailed above.

Then, the scattering problem is solved using the coupled channel Lippmann-Schwinger
equation,
\begin{align} \label{ec:Tonshell}
T_\beta^{\beta'}(z;p',p)=V_\beta^{\beta'}(p',p)+\sum_{\beta''}\int& dp''p''^2
V_{\beta''}^{\beta'}(p',p'')\nonumber\\
&\frac{1}{z-E_{\beta''}(p'')}T_{\beta}^{\beta''}(z;p'',p)
\end{align}
where $\beta$ represents the set of quantum numbers necessary to determine a
partial wave $B\bar{B}JLST$, $V_{\beta}^{\beta'}(p',p)$ is the RGM potential
 and  $E_{\beta''}(p'')$ is the energy for the momentum $p''$
referred to the lower threshold.

The RGM potentials are proportional to the underlying $qq$ interactions, modified with a form factor that encodes the baryon-antibaryon structure. As an illustration, the RGM potentials for the $N\bar N\to Y_c\bar Y_c$ can be written as

\begin{align}
^{\rm RGM} V_{N\bar N\to Y_c\bar Y_c}(p',p)&=  {\cal F}(\vec p^{\,\prime},\vec p)^2 V_{n\bar n\to c\bar c}(\vec{p}^{\,\prime}-\vec{p}).
\end{align}
where  $V_{n\bar n\to c\bar c}$ is the underlying CQM potential and ${\cal F}$ is the $N\to Y_c$ ($\bar N\to \bar Y_c$) form factor,

\begin{align}\label{eq:ffactor}
{\cal F}(\vec p^{\,\prime},\vec p) &= \sqrt{8}\sum_{n_\lambda,n_\lambda'}^{n_{\rm max}}C_{n_\lambda}C_{n_\lambda'}
              \frac{(\eta_{n_\lambda}\eta_{n_\lambda'})^{3/4}}{(\eta_{n_\lambda}+\eta_{n_\lambda'})^{3/2}}
              e^{-\frac{Q^2}{4\left(\eta_{n_\lambda}+\eta_{n_\lambda'}\right)}}
\end{align}
with $Q=\left(\frac{2m_n}{m_c+2m_n}\vec{p}^{\,\prime}-\frac{2}{3}\vec{p}\right)$ the transferred momentum. Here, $C_{n_\lambda}$ ($C_{n_\lambda'}$) and $\eta_{n_\lambda}$ ($\eta_{n_\lambda'}$) are, respectively, the $\lambda$-mode coefficients and ranges of the $N$ ($Y_c$) wave function of Eq.~\eqref{eq:GEMwavef}~\footnote{In this example, the $\lambda$ mode is taken as the coordinate between the heavy quark $c$ and the light pair $nn$. With this choice, the $\rho$ mode (coordinate between the two light $n$ quarks) can be integrated out in Eq.~\eqref{eq:ffactor}.}.

A full coupled-channels study is performed.
The interactions of the diagonal channels, $p\bar p\to p\bar p$ and $Y_c\bar Y_c\to Y_c\bar Y_c$, contribute to the initial and final
state interactions. For the $p\bar p$ channel, the interactions included are the exchange
of $\pi$, $\sigma$ and $\eta$ mesons in the t-channel and the pion and
gluons annihilation in the s-channel. The $Y_c\bar Y_c$ interaction is due to the
$\pi$, $\sigma$ and $\eta$ exchanges in the t-channel and the gluon, $\pi$ and
$\eta$ exchanges in the s-channel. The non-diagonal interaction $p\bar p\to Y_c\bar Y_c$ is
solely due to gluon annihilation in the s-channel.

Before presenting the results, note here that our constituent quark model merges traditional boson exchange and annihilation pictures
in a unified framework. The mesons are generated as Nambu-Goldstone bosons of the spontaneous
chiral symmetry breaking, and gluons as perturbative contributions from QCD.
 For the case of $\Lambda\bar \Lambda$ this mixture allowed for a
description of the total cross section and observables without any fine-tuning of the model parameters.
In the case of single-charmed baryons, the boson exchange diagrams do not contribute to the production
mechanism, so the total cross section will be much smaller than other models based on meson exchanges.

\section{Results} \label{sec:results}

The description of the $p\bar p\to p\bar p$ cross section is a necessary step before describing the single-charmed baryon production.
Previous works~\cite{Entem:2006dt,Ortega:2011zza} focused on the low energy region of the $N\bar N$ system, so we do not expect that the parameters of the optical
potential considered in those works to hold at the $\Lambda_c^+ \Lambda_c^-$ threshold.
For the large production energies of the $Y_c\bar Y_c$ pairs, the optical potential dominates the total, elastic and inelastic $N\bar N$ cross sections.
In order to describe the broad experimental data, gathered by Ref.~\cite{ParticleDataGroup:2020ssz}, we have considered a simple optical potential of the form of Eq.~\eqref{ec:PotOptic}. The parameters are fitted to the $p\bar p\to p\bar p$ total, elastic and inelastic cross sections. A good description of the total, elastic and inelastic cross section is
achieved, as we show in Fig.~\ref{sigppbar}, which validates the model in the relevant energy range. In this figure, a $10\%$ variation of the $V_r$ and $W_i$ parameters of Eq.~\eqref{ec:PotOptic} is included in order to evaluate the sensitivity of the $p\bar p\to p\bar p$ cross sections. Albeit, the impact of such variation in the $p\bar p\to Y_c\bar Y_c$ cross section was found to be negligible.

Due to the lack of experimental data for the $\Lambda_c^+\Lambda_c^-$, $\Lambda_c^+ \Sigma_c^-$ or $\Sigma_c^+ \Sigma_c^-$ channels, it is difficult to
determine the optical potential parameters for the final channels. For the strangeness production~\cite{GarciaOrtega:2012lgi} the $p\bar p\to\Lambda\bar\Lambda$ total cross section was used to fit these parameters but, in this case, there is no available experimental data for the $p\bar p\to Y_c\bar Y_c$.
For this reason, no optical potential will be considered for the final channels. Nevertheless, we have tested the sensitivity of the
$p\bar p\to Y_c\bar Y_c$ total cross section to the optical potential of the final channel using the parametrization for the $\Lambda\bar \Lambda$, $\Lambda\bar \Sigma$+h.c. and $\Sigma\bar\Sigma$ channels used in Ref.~\cite{GarciaOrtega:2012lgi}, comparing them with the results without the optical potential, and we found that the effect of the optical potential is small in the region close to threshold.

\begin{figure}[t]
\includegraphics[width=.45\textwidth]{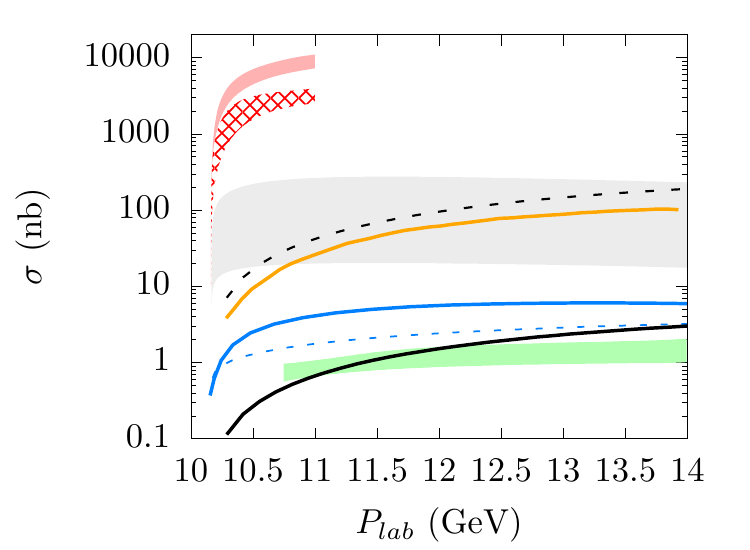}
\caption{\label{sigLLbar} Total cross section for $p\bar p\to \Lambda_c^+ \Lambda_c^-$
for model $(a)$ (black-dashed line) and model $(b)$ (black-solid line), compared
with other studies: The meson-exchange frameworks of Ref.~\cite{Haidenbauer:2016pva} (where solid-red band for meson-exchange and square-red band for quark-gluon transition potential) and Ref.~\cite{Sangkhakrit:2020wyi} (with Regge approach in blue-solid line and Effective Lagrangian in blue-dashed line),
the quark-gluon string (QGS) models of Ref.~\cite{Khodjamirian:2011sp} (grey band) and Ref.~\cite{Kaidalov:1994mda} (orange line) and the perturbative QCD calculation of Ref.~\cite{Goritschnig:2009sq} (green band).}
\end{figure}

We show in Fig.~\ref{sigLLbar} the results for the $p\bar p\to \Lambda_c^+ \Lambda_c^-$ total cross section predicted in this work, compared with other theoretical estimations. We show the total cross section for the two models discussed in Section~\ref{sec:model}, which consider different assumptions for the one-gluon annihilation process $n\bar n\to c\bar c$. The so-called model $(a)$, which takes $m_{ij}=m_n^2$ in Eq.~\eqref{eq:OGEani}, is almost two orders of magnitude larger than model $(b)$, which takes $m_{ij}=\overline{m_\Lambda}^2$, where $\overline{m_\Lambda}$ is the promediated constituent mass of the $\Lambda_c$ baryon, that is $\overline{m_\Lambda}=\frac{1}{3}(2m_n+m_c)$. That indicates the sensitivity of the results to the specific quark propagator in the transition potential. Model $(a)$ is of the same order of magnitude as other theoretical approaches such as the quark-gluon string (QGS) models of Refs.~\cite{Khodjamirian:2011sp,Kaidalov:1994mda}.
On the other hand, model $(b)$, which considers a quark propagator more in line with the energy region of the charmed baryon thresholds, predicts a similar cross section than the perturbative QCD calculation of Ref.~\cite{Goritschnig:2009sq}. In our model, the $n\bar n\to c\bar c$ transition is beyond the chiral symmetry-breaking scale, so no meson exchange is allowed, which explains why the total cross section is much smaller than the estimates from $D$ and $D^*$ exchange frameworks of Ref.~\cite{Haidenbauer:2016pva}. The latter work also considered a $^3S_1$ transition interaction mediated by a gluon, but it is likely that the quark-gluon coupling strength was overestimated. Recent studies of Ref.~\cite{Sangkhakrit:2020wyi} using boson-exchanges mediated by the $D$+$D^*$ mesons with a Regge approach and Effective Lagrangian show smaller cross sections, comparable with our calculations for model $(b)$.

\begin{figure}[t]
\includegraphics[width=.45\textwidth]{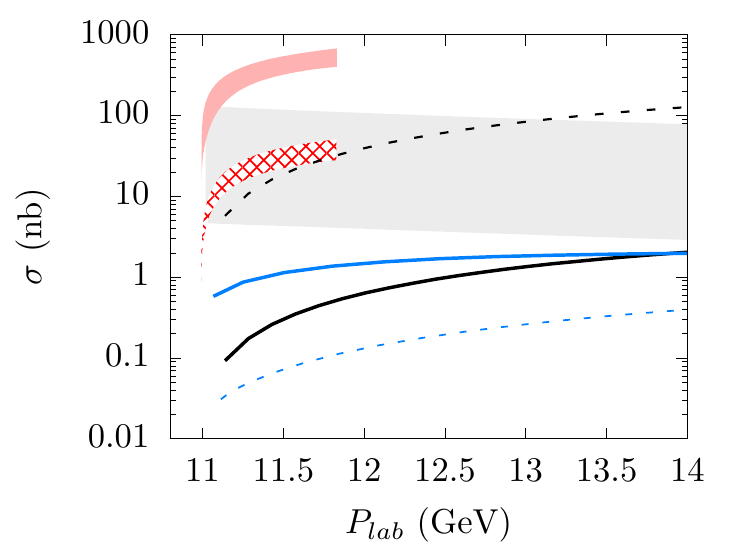}
\caption{\label{sigLSbar} Total cross section for $p\bar p\to \Lambda_c^+\Sigma_c^-$ + h.c. Same legend as in Fig.~\ref{sigLLbar}.}
\end{figure}

\begin{figure}[t]
\includegraphics[width=.45\textwidth]{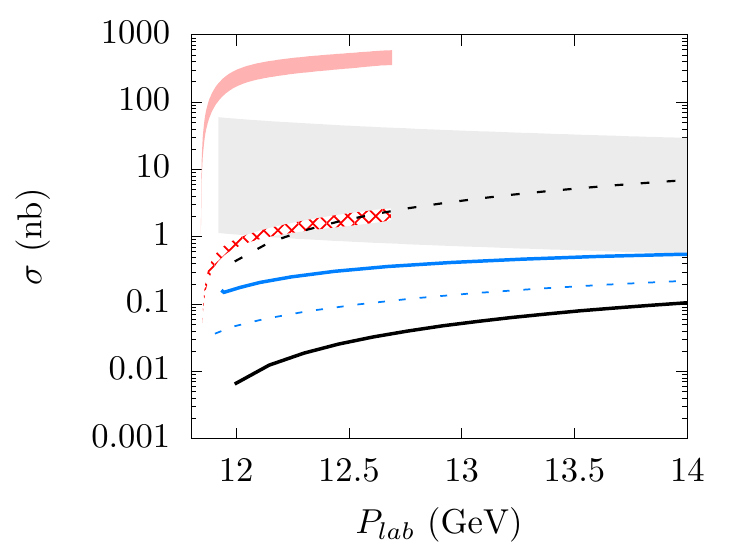}
\caption{\label{sigSSbar} Total cross section for $p\bar p\to \Sigma_c^+\Sigma_c^-$. Same legend as in Fig.~\ref{sigLLbar}. }
\end{figure}

The $p\bar p\to \Lambda_c^+\Sigma_c^-$+h.c., Fig.~\ref{sigLSbar}, shows a similar pattern than the $p\bar p\to \Lambda_c^+\Lambda_c^-$. Interestingly, in this case the total cross section of model $(a)$ is close to the quark model predictions of J. Haidenbauer \emph{et al.}~\cite{Haidenbauer:2016pva} and the QGS of Khodjamirian \emph{et al.}~\cite{Khodjamirian:2011sp}. Model $(b)$, on the contrary, is almost two orders of magnitude below, closer to the results of Ref.~\cite{Sangkhakrit:2020wyi}. Equivalent conclusions
can be obtained for the $p\bar p\to\Sigma_c^+ \Sigma_c^-$, shown in Fig.~\ref{sigSSbar}, but in this case the total cross section is one order of magnitude smaller than those of $p\bar p\to\Lambda_c^+ \Lambda_c^-$ and $\Lambda_c^+\Sigma_c^-$+h.c. for both models.

\begin{figure}[t]
\includegraphics[width=.45\textwidth]{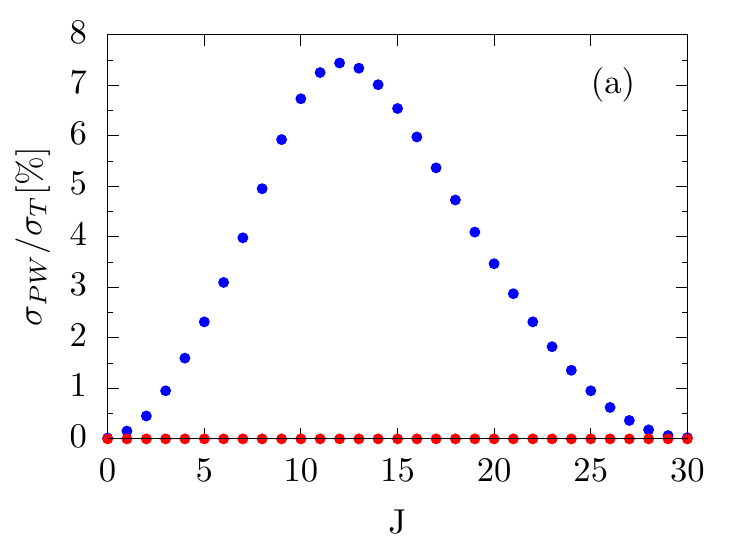}
\includegraphics[width=.45\textwidth]{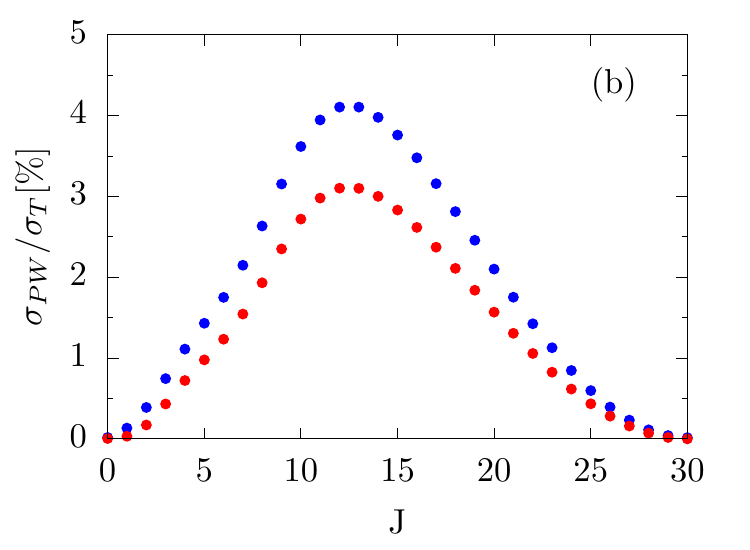}
\caption{\label{sigPW} Contribution of different partial waves, in \%, to the total cross section of $p\bar p\to \Lambda_c^+\Lambda_c^-$ (a) and $p\bar p\to \Lambda_c^+ \Sigma_c^-$+h.c. (b) at $\sqrt{s}=12$ GeV.
Blue dots represent the spin-triplet and red dots the spin-singlet transitions.
}
\end{figure}

In Fig.~\ref{sigPW} we show the contribution of each partial wave to the total cross section, in \%, of $p\bar p\to \Lambda_c^+\Lambda_c^-$ and $p\bar p\to\Lambda_c^+\Sigma_c^-$+h.c.\, at $\sqrt{s}=12$ GeV, depending on their total momentum $J$. The contributions are split on two types: the spin-singlet ($^1L_J\to ^1L_J$, with $L=J$) and spin-triplet ($^3L_J\to^3L_J$ with $L=J-1,J,J+1$) transitions.
We should emphasize that the one gluon annihilation potential is purely central, so the orbital momentum is always conserved and, thus, the non-diagonal $^3(J\pm1)_J\to ^3(J\mp1)_J$ transition interactions are zero.
For both cases, it is interesting to notice a predominance of partial waves with large $J$, with a maximum around $J=12-13$. Such maximum is due to the enhancement of large $J$ partial waves in the initial $N\bar N$ channel for the $Y_c\bar Y_c$ production energy region. The $S$-wave is practically zero, because the $S$-wave of the initial $N\bar N$ is negligible for the energy range under study.

For the isospin-0 $p\bar p\to \Lambda_c^+ \Lambda_c^-$ process (Fig.~\ref{sigPW}$(a)$), the one gluon annihilation potential is zero for the spin-singlet channel, so the only contributions to the cross section are the spin-triplet channels. That is reasonable as the total spin of the $\Lambda_c$ is carried by the charm quark $c$ and the production of the $c\bar c$ is mediated by a vector field, that is, the gluon.
The isospin-1 $p\bar p\to\Lambda_c^+\Sigma_c^-$+h.c. process shows a similar pattern for the spin-triplet channels but, in this case, we have contribution from both spin singlet and triplet channels, with a ratio in the OGE coefficients of Eq.~\eqref{eq:OGEani} of $C_{S=0}/C_{S=1}=3/2$.
The partial wave decomposition for the $p\bar p\to \Sigma_c^+ \Sigma_c^-$ follows the same pattern, with a maximum around $J\sim 12$, having contributions from both spin singlet and triplet components, and also isospin 0 and 1 channels.

This feature produces a abrupt increase of the cross section in the region close to threshold, a behaviour that was also measured in the strangeness production $p\bar p\to \Lambda\bar \Lambda$.

\section{Summary}\label{sec:summary}

In this work we have performed a coupled-channels calculation of the $p\bar p\to \Lambda_c^+ \Lambda_c^-$, $\Lambda_c^+\Sigma_c^-$+h.c. and $\Sigma_c^+\Sigma_c^-$ within a constituent quark
model which properly describes the $N\bar N$ system~\cite{Entem:2006dt} and the production of strangeness in $p\bar p$ collisions~\cite{Ortega:2011zza,GarciaOrtega:2012lgi}.
We have compared the results with other theoretical models, showing that our approach predicts a total cross section consistent with previous studies, in the range of 0.1 to 10 nb for $\Lambda_c^+ \Lambda_c^-$ final channel close to threshold, depending on the assumptions taken in the gluon annihilation transition potential. Considering the energy ranges involved in the reaction, the so-called model $(b)$ seems more realistic, which would lead to a cross section of the order of 1 nb, similar to perturbative QCD estimations~\cite{Goritschnig:2009sq}.

Thus,
this study suggests that the experimental detection of the process of charm production under study can be challenging, but perfectly reachable at $\overline{P}$ANDA considering the planned luminosity of the order of $10^{32}$ cm$^{-2}$s$^{-1}$, which could provide valuable information on the underlying $n\bar n\to c\bar c$ production mechanisms.

\begin{acknowledgments}
This work has been partially funded by Ministerio de Ciencia, Innovación y Universidades under Contract No.~PID2019-105439GB-C22/AEI/10.13039/501100011033,
and by the EU Horizon 2020 research and innovation program, STRONG-2020 project, under grant agreement No. 824093.
\end{acknowledgments}


\bibliographystyle{apsrev4-1}
\bibliography{LcLcpaper}

\end{document}